\documentclass[conference]{IEEEtran}
\IEEEoverridecommandlockouts
\usepackage{cite}
\usepackage{amsmath,amssymb,amsfonts}
\usepackage{algorithm,algorithmic}
\usepackage{graphicx}
\usepackage{textcomp}
\usepackage{xcolor}
\usepackage{url}
\usepackage[a4paper,
            left=0.633in,
            right=0.633in,
            top=0.75in,
            bottom=1.7in]{geometry}
\def\BibTeX{{\rm B\kern-.05em{\sc i\kern-.025em b}\kern-.08em
    T\kern-.1667em\lower.7ex\hbox{E}\kern-.125emX}}
\begin{document}

\title{Deep Back-Filling: a Split Window Technique for Deep Online Cluster Job Scheduling\thanks{This paper has been accepted for presentation at HPCC, 2022. Due to the conference issue, it will be included in HPCC 2023 Proceedings.}}

\author{\IEEEauthorblockN{Lingfei Wang}
\IEEEauthorblockA{
\textit{The University of Melbourne}\\
Melbourne, Australia\\
lingfeiw@student.unimelb.edu.au}
\and
\IEEEauthorblockN{Aaron Harwood\textsuperscript{\textsection}}
\IEEEauthorblockA{
\textit{The University of Melbourne}\\
Melbourne, Australia \\
arnhrwd@gmail.com}
\and
\IEEEauthorblockN{Maria A. Rodriguez}
\IEEEauthorblockA{
\textit{The University of Melbourne}\\
Melbourne, Australia\\
marodriguez@unimelb.edu.au}
}

\maketitle
\begingroup\renewcommand\thefootnote{\textsection}
\footnotetext{Work done while the author was at the University of Melbourne.}
\endgroup
\setlength{\columnsep}{0.24 in}
\begin{abstract}
Job scheduling is a critical component of workload management systems that can significantly influence system performance, e.g., in HPC clusters. The scheduling objectives are often mixed, such as maximizing resource utilization and minimizing job waiting time. An increasing number of researchers are moving from heuristic-based approaches to Deep Reinforcement Learning approaches in order to optimize scheduling objectives. However, the job scheduler's state space is partially observable to a DRL-based agent because the job queue is practically unbounded. The agent's observation of the state space is constant in size since the input size of the neural networks is predefined. All existing solutions to this problem intuitively allow the agent to observe a fixed window size of jobs at the head of the job queue. In our research, we have seen that such an approach can lead to "window staleness" where the window becomes full of jobs that can not be scheduled until the cluster has completed sufficient work. In this paper, we propose a novel general technique that we call \emph{split window}, which allows the agent to observe both the head \emph{and tail} of the queue. With this technique, the agent can observe all arriving jobs at least once, which completely eliminates the window staleness problem. By leveraging the split window, the agent can significantly reduce the average job waiting time and average queue length, alternatively allowing the use of much smaller windows and, therefore, faster training times. We show a range of simulation results using HPC job scheduling trace data that supports the effectiveness of our technique.
\end{abstract}


\section{Introduction}

Online scheduling in High Performance Computing (HPC) centers is of central importance to managers and users of the HPC center alike. Managers tend to prefer that the HPC resources are maximally used, i.e., unused resources only occur if there are no jobs to be scheduled. On the other hand, users prefer short waiting times for their submitted jobs, i.e., the time that their jobs spend waiting in the scheduling queue. However, achieving maximal resource usage usually compromises job waiting times and \textit{vice versa}. e.g., a First-Come-First-Served (FCFS) scheduling policy, which treats all jobs equally in terms of waiting time, may lead to some resources being wasted while the scheduler waits for enough resources to become available for the current first job in the queue. The use of \emph{back-filling} is a widely used heuristic that takes lower-priority jobs from the queue and schedules them to use otherwise wasted resources. However, such jobs may not always exist in the queue, which can lead to 100\% utilization. Some scheduling policies may allow such jobs to wait beyond their earlier possible scheduling time in order to allow for increased utilization, but this can lead to \emph{job starvation} where some jobs are never scheduled. In fact, a naive scheduling policy can easily lead to job starvation, e.g., in a highly loaded system with a high arrival rate of small jobs such that the small jobs are always available to fill resources (achieving maximum utilization) at the expense of unbounded waiting times for larger jobs. FCFS does not suffer from job starvation, but some fundamental scheduling policies such as Shortest-Job-First (SJF) (which is more suited to batch scheduling, i.e., offline) can, and require, additional heuristics to ensure starvation does not occur in an online scheduling system.

Recently, HPC researchers have turned their attention to Deep Reinforcement Learning, where a DRL Agent learns the scheduling policy rather than using a heuristic scheduling policy. The use of DRL is expected to outperform state-of-the-art scheduling heuristics because a DRL Agent has the potential to learn and exploit otherwise unknown or transiently arising patterns of behavior in the underlying HPC system and the user's job submission characteristics. Existing research results show the potential for a DRL Agent to outperform existing heuristics, and they share the following basic approach: an Agent observes a \emph{fixed window of jobs at the head of the queue}, e.g., the first 128 jobs, and sequentially selects individual jobs from the window to be scheduled. However, the following question arises: what is the best way to incorporate back-filling, i.e., when the queue length is greater than the window size?

\subsection{Back-filling}

Some authors switch from the scheduling Agent to a back-filling heuristic when the Agent cannot allocate its selected job~\cite{zhang2020rlscheduler}. This has the advantage of being able to scan the entire queue for possible jobs to back-fill, but it abandons a pure DRL approach which we find undesirable as it may limit the scope for learning. More recently, some authors instead switched from the scheduling Agent to a back-filling Agent that is independently trained to back-fill~\cite{fan2021deep}. This Agent similarly sees only a window of jobs from the head of the queue. When using multiple Agents, a common challenge is that, for either Agent to reach an optimal policy, it needs to learn about the other Agent's actions implicitly. In our approach, we maintain a single Agent, and we avoid the use of heuristics for back-filing. We train our Agent to effectively do both scheduling and back-filling, where in this paper, we contribute a \emph{split window} technique to improve the Agent's ability to back-fill.

Approaches that do not use a heuristic for back-filling, i.e., that are trying to maintain a complete RL approach, share a similar challenge: the job queue can grow to become arbitrarily large, which forces such RL approaches to accept that the job queue is only \emph{partially observable}. This makes it hard to compete with back-filling heuristics that can readily scan the entire queue regardless of its current length. Current approaches can become ``stuck" if their window is not large enough to offer enough candidates for back-filling. Simply increasing the window length leads to an increase in RL model complexity and may not scale to larger dynamic workloads.

\subsection{Our Contribution}

The solution adopted thus far by all state-of-the-art in the literature is to allow the Agent to observe a window of jobs at the head of the queue. In our approach, we effectively overcome the back-filling challenge by adopting what we call a \emph{split window}: some part of the window covers the head of the queue, and the remaining part of the window covers the \emph{tail} of the queue. Intuitively, this approach can allow the Agent to observe every job as it arrives at the tail of the queue, and thus the Agent never becomes stuck without fresh candidates for back-filling. In this paper, we explore this general technique and show considerable increases in scheduling performance over the case when the window is covering the head of the job queue only.

\section{Split Window Technique}\label{split}

In this section, we define the online job scheduling problem and how our split window technique is applied.

\subsection{Online Job Scheduling}
\label{onlinejobsched_sect}

We begin with an online job scheduling model that is widely adopted in the existing Deep RL approaches. An HPC cluster is modeled as a number of cores, $N$, and a job queue, $Q$. Each job, $j\in\{1,2,\dotsc\}$, submitted to the cluster has a submission time, $s_j$, a requested number of cores, $c_j$, and a requested total run-time, $r_j$. The online job scheduling problem requires selecting jobs from the queue, as or after they arrive, that can be run on the cluster and usually assumes that the job queue is unbounded in length. Job $j$ can be run at time $t$ if the number of available cores on the cluster at time $t$ is at least $c_j$. See Table~\ref{tab:ojsnotation}. The problem is to devise a scheduling policy that selects jobs such that a given cluster management objective is met.
\begin{table}[t]
    \centering
    \caption{Online Job Scheduling Notation}
    \label{tab:ojsnotation}
    \begin{tabular}{cl}\hline\hline
    $N$ & number of cores in the cluster \\
    $\mathcal{Q}(t)$ & the set of jobs on the queue at time $t$  \\
    $j$ & job number in $\{1,2,\dotsc\}$  \\
    $s_j$ & the time that job $j$ was submitted \\
    $c_j$ & the number of cores job $j$ requires \\
    $r_j$ & the requested runtime for job $j$ \\
    $W_j$ & the duration of time job $j$ spent waiting on the job queue \\
    $\mathcal{R}(t)$ & the set of jobs running on the cluster at time $t$ \\
    $L(t)$ & the length of the queue at time $t$ \\
    $D(t)$ & the total job load in the queue at time $t$\\
    $\mathcal{J}(t)$ & the set of jobs that have arrived by time $t$ \\
    $\eta(t)$ & spontaneous cluster utilization at time $t$ \\
    %$\zeta(t)$ & spontaneous queue load at time $t$ \\
    $M$ & length of the window, equals $M_{\text{h}}+M_{\text{t}}$ \\
    $M_{\text{h}}$ & the length of the window at the head of the queue \\
    $M_{\text{t}}$ & the length of the window at the tail of the queue \\
    \hline\hline
    \end{tabular}
\end{table}

We let $\mathcal{R}(t)$ be the set of jobs that are running on the cluster at real-time $t>0$, where it follows that $\sum_{i\in \mathcal{R}(t)}c_i \leq N$. The waiting time $W_j\geq 0$ for job $j$ is the (possibly zero) time that job $j$ spent in the queue. In our work, the agent always has a chance to schedule job $j$ at time $s_j$, i.e., without any queueing delay, since we allow the agent to take a scheduling action at every job arrival time and job completion time, and we serialize the jobs that arrive together, to ensure that the agent can see every job without delay.

\subsubsection{Average Utilization} The fraction of cores used at time $t$ is the spontaneous utilization, $\eta(t)=\frac{| \mathcal{R}(t) |}{N}$, of the cluster. Or we can consider the mean utilization of the cluster at time $t$ over the previous time interval $[t-\Delta,t)$: 
\begin{equation}
    \label{utilization}
\Tilde{\eta}(t;\Delta)=\frac{1}{\Delta}\int_{t-\Delta}^{t}\eta(t)dt.
\end{equation}

\subsubsection{Average Queue Length} The number of jobs waiting in the queue, $L(t)$, at time $t$, and similarly, $\Tilde{L}(t;\Delta)$ over a time interval.

%\subsubsection{Average Queue load} The total spontaneous load waiting on the queue, $\zeta(t)=\sum_{j\in Q(t)}c_j\,r_j$, at time $t$, and similarly, $\Tilde{\zeta}(t;\Delta)$.

\subsubsection{Average Waiting Time} As defined, $W_j$, is the time spent by job $j$ waiting on the queue. In this case, we consider the set of jobs that have arrived by time $t$, $\mathcal{J}(t)=\{j\, \mid \,s_j < t \}$, and consider the average waiting time:
\[
\Tilde{W}(t) = \frac{1}{|\mathcal{J}(t)|}\sum_{j\in \mathcal{J}(t)} W_j 
\]

\subsubsection{Average Job Load} The average amount of job load in the queue. The load for job $j$ is defined as $c_j \times r_j$. Similarly, the average job load in the queue at time $t$ is:
\[
\Tilde{D}(t) = \frac{1}{t}\sum_{j\in \mathcal{J}(t)} W_j \times c_j \times r_j
\]

%Since the starting time for job $j$ is $s_j+W_j$, the amount of time remaining for a job $j\in R(t)$ is at most $r_j-s_j-W_j$.

\subsection{Agent's Window and Window Split}

The agent's window of constant length $M$ can be defined by the set of $M$ indices into the queue that the window covers, which is, in the default case, just the head of the queue: $\{1,2,\dotsc,M\}$. In our work, the window is more generally a subset of $\{1,2,\dotsc,L(t)\}$, where we consider selecting $M_{\text{h}}$ indices $\{1,2,\dotsc,M_{\text{h}}\}$ from the head of the queue and $M_{\text{t}}$ indices $\{L(t),L(t)-1,L(t)-2,\dotsc,L(t)-M_{\text{t}}+1\}$ from the tail of the queue, where $M_{\text{h}}+M_{\text{t}}=M$.

\subsection{Insight}

In our research, we observed that when the agent's window into the job queue covers only the head of the queue, the agent sometimes accumulates jobs in the window that currently cannot be scheduled, and such jobs sit at the head of the queue. If the number of such jobs starts to become equal to the window size, then the agent can reach a state where no jobs in the window can be scheduled. At this point, the agent can only wait for the cluster to complete some jobs so that a job in the window can be placed, possibly revealing one more job from outside of the window. Such a ``wait" state is undesirable because it does not allow the agent to learn how to back-fill, and for queues whose length greatly exceeds the window length, the agent loses opportunities to back-fill.

Somewhat counter-intuitively, as defined earlier, we decided to allow the window to cover some jobs at the tail of the queue so that the agent's \emph{effective window} was a combination of jobs from the head of the queue and jobs from the tail of the queue. While the agent's total window size remains fixed, the agent now has the potential to observe \emph{every} job that arrives on the queue, which completely avoids the ``wait" state discussed above.

\section{Deep Back-filling}

To demonstrate the effectiveness of the split window technique, we developed a job scheduling simulation environment and framework for agent training, similar to existing approaches but with the following two significant differences:
\begin{itemize}
\item \textbf{Schedule Cycling} -- at any given scheduling decision, the agent can choose not to schedule a job (even if a job could be scheduled) but rather just wait until either a running job has been completed or a new job has arrived on the queue.
\item \textbf{Split Window} -- the agent can observe $M_h$ jobs at the head of the queue and $M_t$ jobs at the tail of the queue at each scheduling decision.
\end{itemize}
For convenience, we refer to our overall approach as \textbf{Deep Back-filling} or DBF. In this section, DBF is explained in detail with the RL representations -- state, reward, and action. Next, we present our Schedule Cycling, the training strategy, which defines when and how the scheduling decision is made, which makes it possible for our agent to learn higher-level scheduling strategies.

\subsection{Deep Back-filling Framework}\label{framework2}
\begin{figure}[tbp]
\centerline{\includegraphics[width=90mm,scale=0.5]{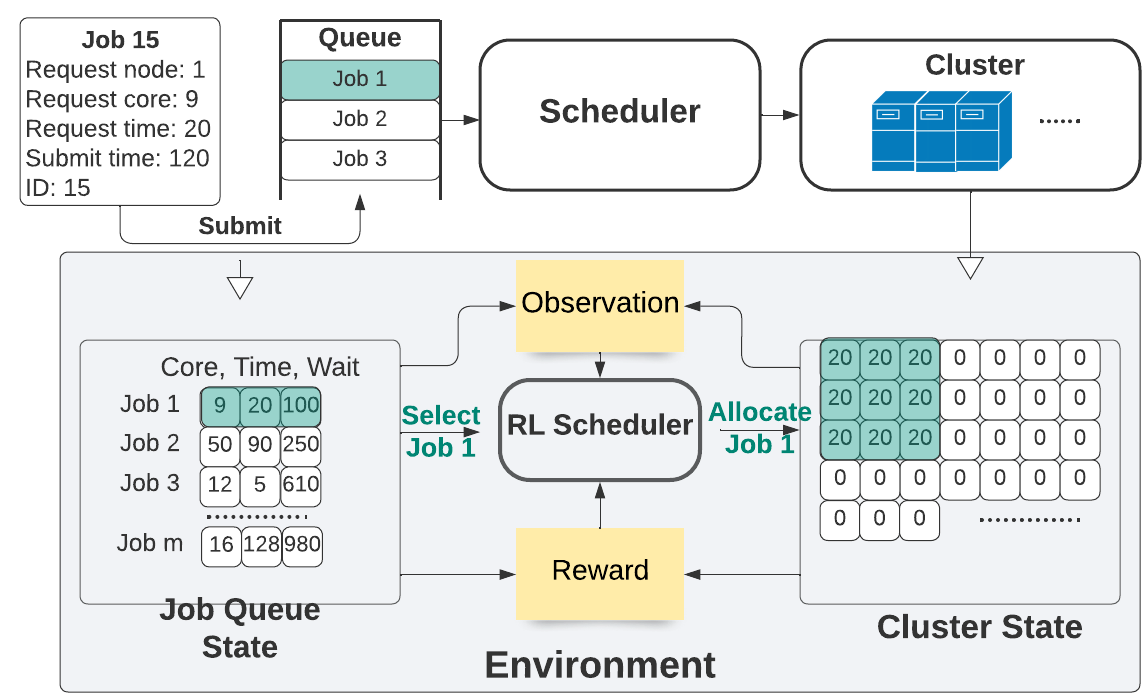}}
\caption{The DBF framework.}
\label{framework}
\end{figure}
Fig.~\ref{framework} shows an overview of the DBF framework. In the upper part of the figure, users submit their jobs to the job queue by specifying the resources needed. The scheduler makes scheduling decisions based on the cluster's status and the job queue. The bottom part shows the RL representations of the scheduling problem.

\subsubsection{Observation}

The observation of the RL scheduler is the union of the cluster state and the window of jobs from the job queue state. Cluster state, $\sigma(t)=\big[\sigma_1(t)\;\sigma_2(t)\;\cdots\;\sigma_N(t)\big]$, defines the state of each core at time $t$, where 
\[
\sigma_i(t)=\begin{cases}
0 & \text{core $i$ is available}\\
t-s_j-W_j & \text{job $j$ is running on core $i$}
\end{cases}
\]
When job $j=<c_j, r_j, W_j(t)>$ is running on core $i$ then $\sigma_i(t)$ is the time remaining for core $i$ to become available. Here, $W_j(t)=t-s_j$ represents the waiting time of job $j$ in the queue at time $t$.

\subsubsection{Action} The set of actions, $\mathfrak{A}=\{1,2,\dotsc,M\}\cup\{\text{Fwd}\}$, includes selecting one job from $M$ jobs in the window, and \emph{Fwd} which represents a ``forward" action where the agent can wait for either a job on the cluster to complete running or another job to arrive before selection a job from the window, as described further in Section~\ref{cycle}.  

%The \textbf{action} of the RL agent is either choosing one job from the $M$ jobs or selecting the \textbf{Forward} action. The details of forward action are explained in the sections \ref{cycle}.

%If the element is zero, it means the core is free otherwise it is executing a job with a certain wall time. 

\subsubsection{Reward} The reward function consists of the current unused resource penalty, current queue length, and current total waiting time of jobs in the queue. Instead of optimizing a single objective, we maximize the whole system's performance by considering several objectives simultaneously. At time $t$, we use a reward:
\begin{equation}\label{reward}
\mathfrak{R}(t) =
%\begin{cases}
-w_1\,(1-\eta(t))-w_2\,\frac{L(t)}{L_{\text{max}}(t)}-w_3\,\frac{W(t)}{W_{\text{max}}(t)}
%0, \text{when the agent place the job}
%\end{cases}
\end{equation}
where
\begin{align*}
W(t) &= \sum_{j \in \mathcal{Q}(t)}(t-s_j) \\
L_{\text{max}}(t) &= \max_{\tau\in [0,t]}\big\{L(\tau)\big\}\\
W_{\text{max}}(t) &= \max_{\tau\in [0,t]}\big\{W(\tau)\big\}
\end{align*}
and $0\leq w_1,w_2,w_3\leq 1$ are constants of proportionality.

\subsection{Schedule Cycling and Rewards}\label{cycle}
\begin{figure}[tbp]
\centerline{\includegraphics[width=90mm,scale=0.5]{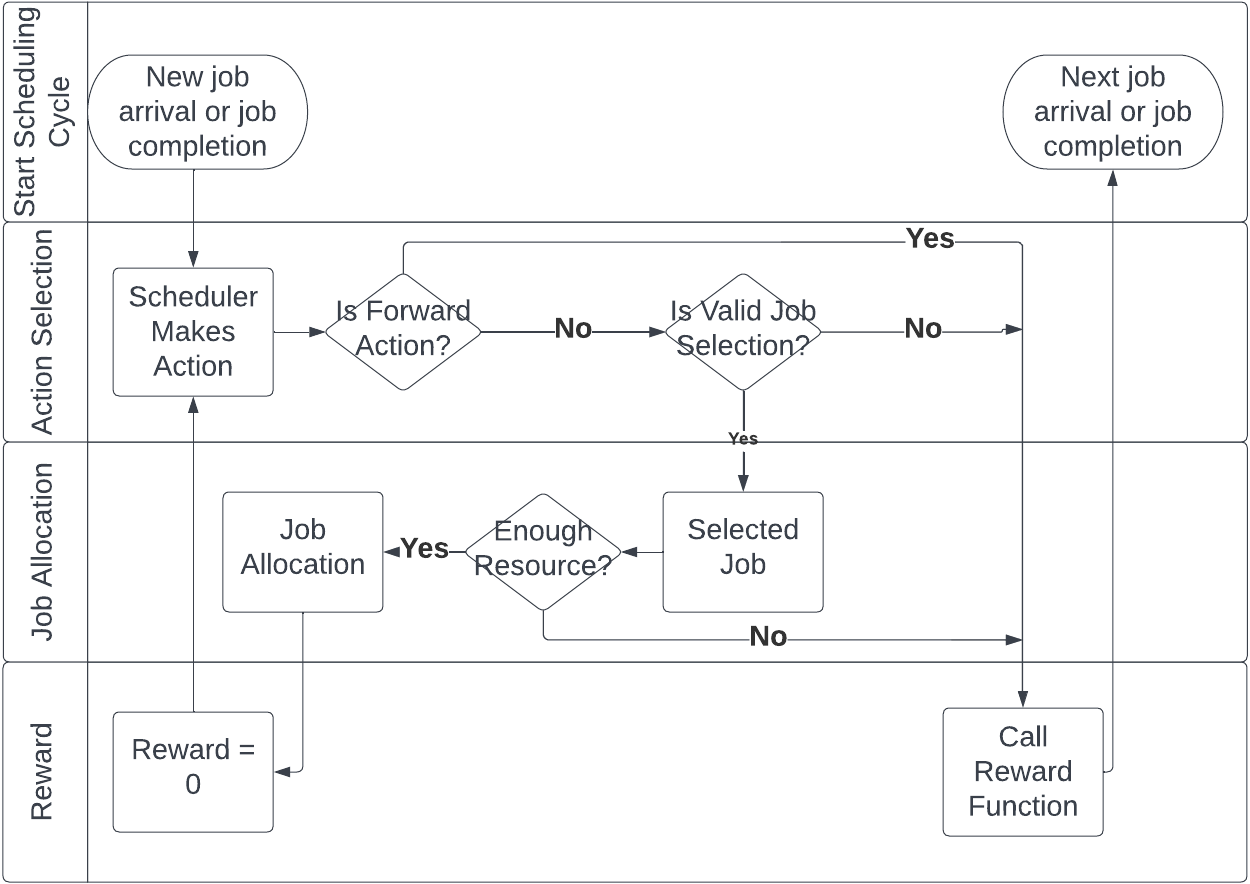}}
\caption{The flowchart of Schedule Cycling.}
\label{flowchart}
\end{figure}
In general, to train a reinforcement learning agent, the agent either receives the rewards immediately after the actions or receives a delayed reward when it reaches special states or rules. The agent can learn quickly for an immediate reward by evaluating its action step-by-step. However, the training may end up with a suboptimal policy when the agent tries to maximize the reward for each step and loses the "global vision" of the problem. For the delayed rewards, the agent can overview the problem and learn to achieve an overall better result. But it can fail to learn when the delay is too long.

For the existing works, such as RLScheduler~\cite{zhang2020rlscheduler}, SchedInspector~\cite{SchedInspector}, and DARS~\cite{fan2021deep}, if a job is selected by the agent when the cluster does not have sufficient resource to run the job, their schedulers will wait for other job completions and apply back-filling until the job can be run. Pathological cases where the selected job is large can significantly degrade the agent's performance since it may have to wait for some time before making any decisions.

To solve the problem, we designed a new scheduling mechanism -- Schedule Cycling, shown in Fig~\ref{flowchart}. A scheduling cycle is triggered at time $t$ when a new job $j$ arrives on the queue at time $t=s_j$ or a job $j$ completes running on the cluster at time $W_j+r_j$. The agent receives the observation $\sigma(t)$ and selects an action from $\mathfrak{A}$. The Fwd action will terminate the current scheduling cycle, receive a negative reward $\mathfrak{R}(t)$, and move the system forward in time to the next scheduling cycle (next new job arrival or running job completion). If the selected action is in $\{1,2,\dots,M\}$, we check whether it is valid or not. An invalid selection arises when the scheduler chooses the $i^{th}$ job from the window and either $L(t)<i$ or $N-|\mathcal{R}(t)|<c_i$. For invalid selections, the system will be forwarded to the next scheduling cycle and receive a negative reward $\mathfrak{R}(t)$. If the selection is valid, then the job is removed from the queue and allocated to the cluster with a zero (maximum) reward given to the agent, and the agent needs to continue making scheduling decisions until the system is forwarded. In this way, the Schedule Cycling mechanism allows the agent to learn to place jobs to maximize the reward, and the agent can also learn to achieve better job placement with the Fwd action when a more suitable job arrives in the near future.
%We assume the time is frozen within the Schedule Cycling because each scheduling decision made by our scheduler only takes about $0.4 ms$. However, the request time of the job spread from hours to months in HPC.

Using Schedule Cycling, our scheduler not only leverages the delayed reward in a manageable number of transitions so as to take an overview of the problem but also has a chance to learn whether to wait or not, as only valid selections will be placed. The overall strategy causes the scheduler to learn to make valid job allocations without an explicit penalty. Furthermore, including the waiting time in the observation and in the reward function, $\mathfrak{R}(t)$, prevents job starvation. Therefore, our DBF framework can learn a more advanced policy compared to without Schedule Cycling and does not require a separate back-filling heuristic nor a back-filling agent.

\section{Experimental Results}
%\subsection{Implementation}
\subsection{Simulation and Agent Implementation}

We simulate an HPC cluster as described in section~\ref{onlinejobsched_sect}. Jobs and job inter-arrival times are modeled based on commonly used traces, specifically the Lublin dataset~\cite{lublin2003workload}. The Agent's PPO algorithm is implemented using PyTorch~\cite{paszke2019pytorch}. In DBF, training occurs in an episodic manner, with an episode ending when 1,000 jobs have been successfully placed, whether there are remaining jobs in the queue or not. Since the scheduler can make invalid job selections, the length of each episode is not fixed. After an episode terminates, an update to the actor and critic networks is performed by stochastic gradient descent with a mini-batch size of 128. The learning rate $\alpha$ is $3\times10^{-4}$, clip $\epsilon$ is $0.2$ and the discount factor $\gamma$ is $0.99$. The hyperparameters for the reward function are $w_1=w_2=w_3=\frac{1}{3}$. The hidden layer size for actor and critic networks is $[1024, 512, 256]$. The listed hyperparameters come from a hyperparameter search.

We normalize the Agent's observation, $\bar{\sigma}(t)=\tfrac{1}{r_{\text{max}}\sigma(t)}$, where $r_{\text{max}}=\max_{j\in \mathcal{J}}\{r_j\}$ and $\mathcal{J}$ is the set of jobs arriving overall time, as training with normalized states has been shown to speed up learning in RL problems~\cite{andrychowicz2020matters}. In particular, we use the maximum allowed runtime of the system to normalize the cluster's state vector and the maximum number of processors, maximum allowed runtime, and maximum waiting time of all jobs to normalize the state vectors of queued jobs.

\subsection{Evaluation Setup}
As previously mentioned, we use the Lublin workload model~\cite{lublin2003workload} to evaluate our approach. The reason is twofold: \emph{i}) it is widely used in the related literature, and \emph{ii}) it represents a realistic HPC workload as it is based on execution traces of thousands of jobs spanning several months. Specifically, the trace data contains all jobs that arrived in 89 days on a 256-core cluster. Each job requests 1 to 256 cores, $c_j \in [1,256]$ and 1 second to 34.6 hours runtime, $r_j \in [1, 124707]$. The average requested core-time per job ($r_j \times c_j$) is $209,278$ seconds, and the average amount of core-time submitted to the cluster is $271.5$ per second. To avoid over-fitting -- which may arise if the same sequence of jobs is used across episodes -- we use a sequence of jobs starting at a random point in the workload for each episode.

We compare the performance of DBF to three heuristics, as they are widely used in real-world systems, and one RL-based scheduler, the newest work in the related literature:
\subsubsection{First-Come-First-Served} (\textbf{FCFS}) schedules the jobs in the order of their arriving time. FCFS with back-filling is the default scheduler for many HPC clusters and works reasonably well in practice. 

\subsubsection{Shortest-Job-First} (\textbf{SJF}) gives higher priority to smaller jobs and thereby always places the job with the smallest number of requested cores.

\subsubsection{Last-Come-First-Serve} (\textbf{LCFS}) schedules jobs with the latest arrival first.
LCFS can suffer of starvation since the probability of scheduling the job at the head of the queue is low.

\subsubsection{SchedInspector} is an RL-based HPC scheduler~\cite{SchedInspector} proposed by Zhang, Dai, et~al. The main idea behind SchedInspector is to train an RL agent to determine whether to accept or ignore the job placement choice made by a heuristic-based scheduler. The authors hypothesize that ignoring certain scheduling actions in some states can lead to better overall performance. We implemented their approach by using the code released in their \emph{github} repository\footnote{\url{https://github.com/DIR-LAB/SchedInspector}} with the default objective of minimizing the average bounded job slowdown (the average fraction of the job waiting time plus job execution time to the job waiting time) and back-filling enabled. We use both Shortest Job First (SJF) and F1~\cite{carastan2017obtaining} as the base heuristic schedulers, as also used by Zhang, Dai, et al.

SchedInspector and DBF use different training methods. SchedInspector trains on a job sequence of 128, and it stops receiving new jobs when 128 jobs have been submitted to the system. The episodes end after the 128 jobs have been scheduled, no matter how long it takes. We call it offline training. However, an online system can always receive new jobs during operation, and we use this way to train our DBF. To make a fair comparison, we will present the results of SchedInspector in both online and offline testing methods in section~\ref{result_comparison}.

%\textbf{RLScheduler}~\cite{zhang2020rlscheduler} is an RL-based scheduler whose reward function is average job slowdown. The RLScheduler only takes jobs as the state, which contains job attributes and the earliest running time of jobs. It has no information about the cluster. The action is selecting a job from a bounded queue. Instead of discarding the invalid action, the RLScheduler will wait until the selected job can be run before making the next action.

\subsection{DBF Performance and Schedule Cycling Validation}\label{result_comparison}
\begin{figure}[htbp]
\centerline{\includegraphics[width=80mm,scale=0.5]{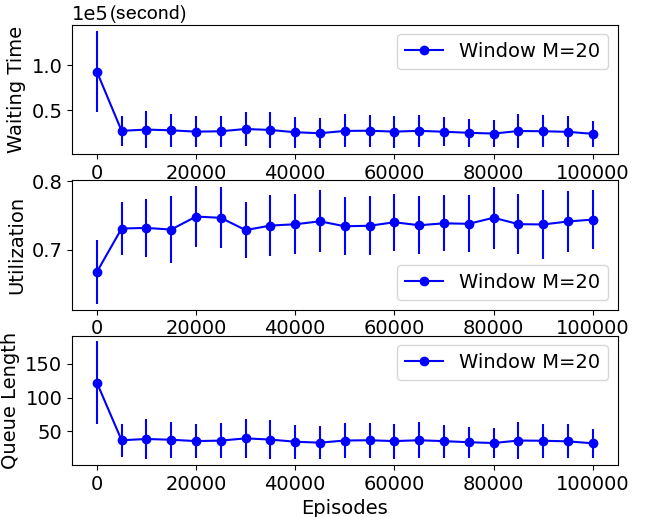}}
\caption{The learning curves of DBF with a window size of 20 for three different objectives.}
\label{learning}
\end{figure}

We first evaluate the effectiveness and efficiency of DBF when learning to schedule jobs and optimizing the system's objectives (minimizing queue length and waiting time and maximizing utilization) in an online environment. We test a range of window size configurations. We trained the agent for 100,000 episodes and saved trained models after every 5,000 episodes. Then, we evaluated each of the saved models for 100 episodes without updating. Fig.~\ref{learning} plots the learning curves of DBF with a window size of 20 without a split window, $M=20$ and $M_t=0$; similarly, the results for window sizes of 32, 64, and 128 share the same patterns. The points are the average performance for the objectives at each checkpoint, and the error bars are the standard deviations. Fig.~\ref{learning} shows that DBF converges successfully by continuously reducing the average queue length and job waiting times while increasing the utilization of the cluster. After approximately 5,000 episodes, which we consider a reasonable amount of interactions with the environment, the average utilization, waiting time, and queue length improve dramatically. Afterward, the improvements become smaller because the agent cannot observe all jobs in the queue as the window size is 20, where the average queue length is 31.

%converging to a state whecan not only optimize the multiple goals simultaneously, but the learning process is very fast. After only 5000 episodes, the average utilization, waiting time, and queue length have improved dramatically. Then, the improvements become smaller because the agent cannot observe all jobs in the queue where the average queue length is 31.

\begin{figure*}[tbp]
  \includegraphics[width=\textwidth]{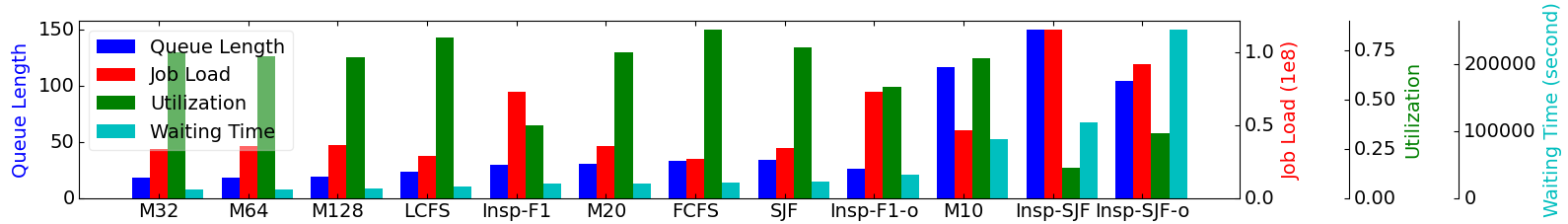}
  \caption{Performance comparison (sorted by average waiting time) on trained DBF of different window sizes with other schedulers. M10 means the DBF has a window size of 10. Insp-F1-o means the SchedInspector method has an F1 algorithm as the base scheduler and the test method is offline. Insp-SJF means testing SchedInspector in an online method with SJF as the base scheduler.}
\label{back-fill}
\end{figure*}

Next, we evaluate whether DBF can learn back-filling by comparing its performance to specific heuristics and RL-based models with back-filling. From Fig.~\ref{back-fill}, we can see the DBF with window sizes of 32, 64, and 128 can reach better average waiting times and queue lengths with a relatively good utilization in all compared algorithms. The heuristics can achieve slightly better average utilization and job load in the queue. The SchedInspector-based approaches lead to the lowest cluster utilization for both online and offline tests. The main reason is that SchedInspector skipped most of the scheduling choices made by its base-heuristic algorithms, leaving the cores idle for long periods. Also, SchedInspector optimizes a single objective, average bounded job slowdown. In terms of average job slowdown, SchedInspector with SJF (average job slowdown 62.2) can outperform the SJF (average job slowdown 73.3). But SchedInspector is much worse than SJF when comparing the direct measurements, such as utilization and average waiting time. In our experiments, we observed such trade-offs between scheduling objectives, which means gains in one objective are achieved at the expense of other ones. For this reason, in DBF, the hyperparameters of each objective are adjustable in the reward function. 

%SchedInspector uses an offline training method, which fixes the number of arriving jobs per episode to a set value, 1000 in this case. DBF, on the other hand, 

%maximizing or minimizing one of the objectives can cause another to lose so that the hyperparameters of each objective are adjustable in DBF. Overall, DBF achieved the best performance with a window size of 32. It suggests that our design of Schedule Cyclings and delayed reward makes it possible for a single agent to learn scheduling and back-filling simultaneously.

\subsection{DBF Split Window}
The pre-defined size of the input limits the performance of the neural network-based scheduler. If the system received jobs more than the window size, the partially-observed state could cause the agent to learn a suboptimal policy. A natural method is to increase the window size. However, continuously increasing the window size can make the training more challenging and does not guarantee an increase in performance. According to the discussion in Section~\ref{split}, the split window technique can be a potential solution for the problem as it has a chance to observe all newly arrived jobs in the tail of the queue at least once. The agent gets at least one opportunity to schedule all jobs.
In this section, we compare the performance of DBF with and without the split window technique to evaluate the method.

Fig.~\ref{split} plots the different splitting configurations with each window size. The titles of the subplots show the total window size, and the x-axis indicates the tail size. It shows the performance of each model after training for 100,000 episodes. The left y-axis is the average waiting time, and the right y-axis is the average queue length. The utilization is not plotted as all models share a similar utilization. With the window sizes of 8, 10, 16, and 20, we can clearly see that increasing the tail window size helps to improve the performance. The performance is significantly enhanced when it includes just one job from the tail with smaller total window sizes. It reveals that the split technique works extremely well, especially with smaller window sizes. It also suggests that we can achieve the same level of performance with less information by using the splitting. Moreover, agents with smaller window sizes are much easier to train, so it also improves sample efficiency. From the subplots of window sizes 32 and 64, the improvement of splitting is not as significant as in the smaller window sizes. Because 32 and 64 are larger than the average queue length. These agents can observe all jobs most of the time, but splitting windows can also improve performance. Unlike window sizes of 8, 10, and 16, continuing to increase cannot always have a positive impact on window sizes of 20, 32, and 64 as the performance gets worse when taking too many jobs from the tail.

\begin{figure}[htbp]
\centerline{\includegraphics[width=80mm,scale=0.5]{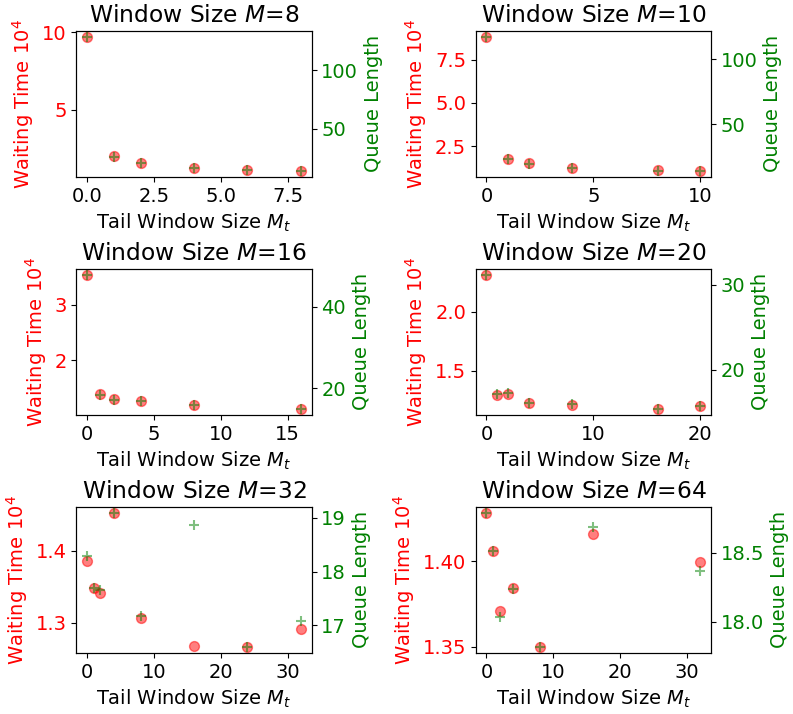}}
\caption{Performance comparison on different window sizes with various splitting configurations. The red points represent the average waiting time, and the green crosses represent the average queue length.}
\label{split}
\end{figure}

\begin{figure}[tbp]
\centerline{\includegraphics[width=90mm,scale=0.5]{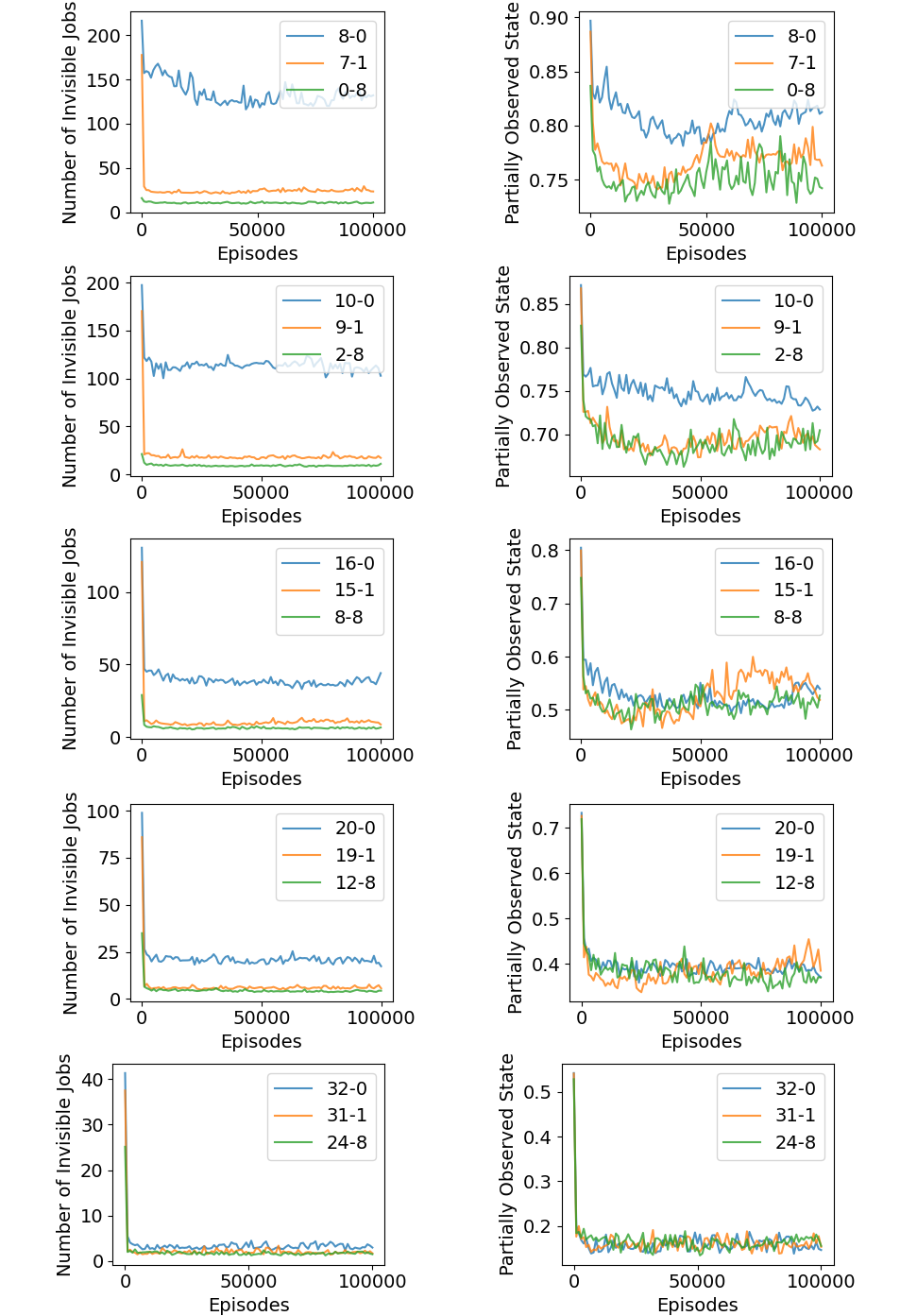}}
\caption{The average number of invisible jobs (cannot be observed by the agent) and the average ratio that the observation is partially observed in the training. The indicated window configuration $M_\text{h}$-$M_\text{t}$ means the head window size is $M_\text{h}$ while the tail window size is $M_\text{t}$.}
\label{exceed}
\end{figure}

Fig.~\ref{exceed} provides insights into how the splitting improves the average queue length in training. The plots on the left show the number of invisible jobs (how much the queue length exceeds the window size) on average. The plots on the right give information about the percentage of the partially observed state that the queue length exceeds the window size. For window sizes of 8 and 10, the splitting window can significantly decrease the average exceeding length and the ratio. The exceeding length can be largely decreased for window sizes 16 and 20 while the ratios remain at the same level using the splitting technique. When the window size is large compared to the average queue length, the improvement is limited in terms of average exceeding length for window size 32.

\begin{figure}[htbp]
\centerline{\includegraphics[width=90mm,scale=0.5]{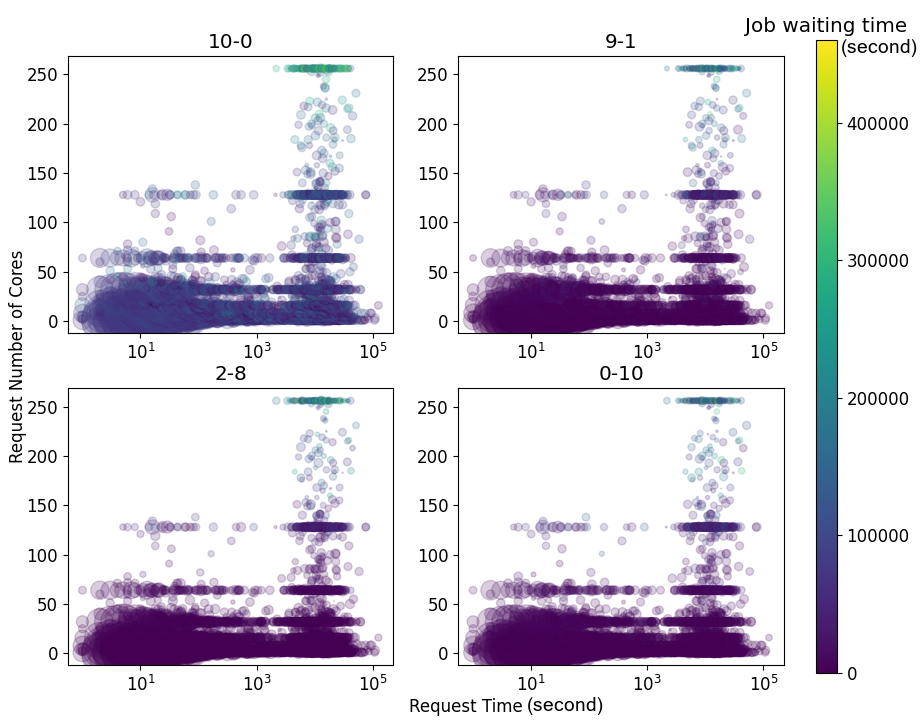}}
\caption{The average waiting time and the number of times being placed for each job type in 1,000-episode testing. The color indicates the average waiting time, and the circle size indicates the number of times the type of job is placed (a bigger circle means the job is placed more frequently). The titles indicate the window configuration $M_\text{h}$-$M_\text{t}$ that means the head window size is $M_\text{h}$ while the tail window size is $M_\text{t}$.}
\label{starve}
\end{figure}

Moreover, Fig.~\ref{starve} illustrates how the split technique improves the average waiting time for each type of job using the window size of 10 as an example. We use the requested cores and runtime to distinguish job types which are shown on the x-axis and y-axis. Every circle in the plot represents a type of job, while the average waiting time is shown in the colors, and the circle size is the number of times that type of job is placed. It is clearly demonstrated in Fig.~\ref{starve} that even one split window at the tail of the queue can dramatically decrease the average waiting time for nearly every type of job. However, as discussed, if the tail window size is closer to the total window size, the risk of big jobs at the head of the queue being starved is higher. When comparing the sizes of the circles in the upper area in each subplot, the 0-10 configuration has smaller sizes for big jobs (the requested number of cores is greater than 128). It suggests that if the agent only takes information from the tail of the queue, the big jobs at the head of the queue are invisible to the agent so that the jobs cannot be starved. Although the configuration of 0-10 achieves the best average waiting time, such a window configuration should not be used in the real system to avoid job starvation.

%As discussed, if the tail window size closes to the total window size, the risk could be higher that the big jobs at the head of the queue being starved. We take the window size of 10 as an example to analyze the potential job starvation problem. Fig.~\ref{starve} illustrates the average waiting time and the frequency of placement of the jobs requested 256 and 128 cores in window-size-10 configurations. The reason for choosing 256-core and 128-core jobs is that the types of jobs request relatively more resources, which are most likely to be starved. The 10-0 model has the longest average waiting time for the 256-core jobs and the biggest circle size among 256-core jobs. Without the splitting, the agent only sees the head of ten jobs so that it can place more big jobs in the head but with the longest waiting time. For configurations of 9-1, 8-2, 6-4, and 2-8, the sizes of the 256-core-job circles are similar because all of them observe the head of the queue and have chances to allocate the big jobs. However, the model of 9-1 has a relatively larger waiting time. The 0-10 model has the smallest circle for 256-core jobs. It suggests that if the agent only takes information from the tail of the queue, the big jobs at the head of the queue have a higher chance to be starved. For 128-core jobs, as the 10-0 agent only places the jobs at the head of the queue, it schedules more big jobs than others with a slightly smaller waiting time, since the big jobs are easier to wait at the head of the queue. And others share a similar frequency and waiting time.

In conclusion, the splitting technique can increase performance, especially when the window size is smaller than the average queue length. Using the splitting window, the agent can achieve the same level of performance with less information (window size). However, increasing the splitting does not always have positive impacts. When the window size and tail window size are big enough, continuously expanding the tail window size may cause a performance loss. Although the splitting seems to improve performance, it may lead to job starvation. Thus, the agent should not take all observations from the tail of the queue.

\section{Related Work}

Although recent developments in RL-based scheduling have shown promising results, RL-based schedulers have not been sufficiently studied in HPC contexts. RLScheduler~\cite{zhang2020rlscheduler} utilized Proximal Policy Optimization (PPO)~\cite{schulman2017proximal} to build a batch job scheduler for HPC systems. The results demonstrate that the DRL-based scheduler outperforms heuristic-based schedulers for some workloads when considering performance in terms of job slowdown or job waiting time. However, it only shows a limited improvement compared to the best-performing heuristic. RLScheduler doesn't support back-filling actions; it relies on a simple back-filling heuristic to fill the system's holes with possible jobs. Also, the Agent optimizes a single scheduling objective, either average job slowdown or average waiting time. 

SchedInspector~\cite{SchedInspector} builds an inspector based on reinforcement learning. The inspector takes into consideration the cluster's and queue's state to determine whether to execute or ignore the scheduling decision made by a heuristic algorithm. A scheduling decision may be ignored with the aim of making better scheduling decisions in the future, which can improve the scheduling objective. However, the performance of this method largely depends on the base heuristic. Further, it relies on a separate heuristic-based back-filling algorithm.

DRAS, proposed by Fan et al.~\cite{fan2021deep}, uses two DRL agents, one to perform job selection and another one to make back-filling decisions. The two agents observe the job queue with a window size of 50. The authors evaluated DRAS with two reinforcement learning algorithms, Deep Q-Network (DQN)~\cite{mnih2013playing} and Policy Gradient (PG)~\cite{sutton1999policy}. The results suggested that DRAS with PG achieves an overall best performance when compared to DRAS with DQN, Decima~\cite{mao2019learning}, and other heuristics, such as FCFS, BinPacking. For DRAS, more than 83\% of the jobs are back-filled, which means the back-filling agent is used more frequently. During the training process, DRAS used the same workload in every episode, which can lead to overfitting the scheduler to the training workload.

Based on PPO, Wang et al.~\cite{wang2021rlschert} designed a job scheduler named RLSchert. The key improvement of RLSchert is that it enables a prediction model to estimate the remaining runtime of jobs by feeding a recurrent neural network-based model with the features derived from HPC jobs. The experimental results showed that their scheduler outperformed DeepRM~\cite{mao2016resource} and commonly used heuristics in terms of average slowdown and resource utilization. However, inaccurate runtime predictions can lead to jobs being killed by the scheduler, resulting in resource wastage and reduced user satisfaction. As with DRAS, the model is trained using the same workload in every episode.

Orhean et al.~\cite{orhean2018new} examined the scheduling problem in heterogeneous distributed systems with Q-Learning and State–Action–Reward–State–Action (SARSA) algorithms. The results show that Q-Learning and SARSA worked for a two-node environment but failed when increasing the number of nodes.

Instead of choosing a job to schedule, CuSH~\cite{domeniconi2019cush} presented a two-step RL-based solution for HPC job scheduling. Firstly, CuSH uses a job selector to select a job from a visible job queue, then a policy selector can select one of the two proposed policies to place the job. The job selector and policy selector are both based on Q-Learning, but they are updated separately. By separately training the agents, the system may lack an overview of the whole problem making it hard to achieve overall maximum rewards.

\section{Conclusion and Future Work}
A reinforcement learning-based batch job scheduler, DBF, is presented in this study. Unlike other RL-based schedulers, which rely on a separate back-filling scheduler, DBF can learn how to back-fill with the novel Schedule Cycling mechanism. The results, run in an online environment, show that our approach outperformed the schedulers with a back-filling heuristic regarding the average job waiting time and queue length. We design and evaluate a new technique called the split window. With the splitting, the scheduler can observe the jobs at the head of the queue and get a chance to place all newly arrived jobs at the tail of the queue. The results demonstrate a significant improvement when applying the technique. When the queue size is small compared to the average queue length, the splitting can reduce the average waiting time and queue length to a different magnitude. Even with a relatively large window size of 20, the splitting can reduce the average waiting time and queue length by 49\% and 50\%. However, increasing the splitting size cannot always help. Therefore, finding the optimal splitting configuration by leveraging machine learning techniques can be the future work of this study.
%\section*{Acknowledgment}

\bibliographystyle{ieeetran}
\bibliography{Reference.bib}

\end{document}